\documentclass{article}
\usepackage{spconf,amsmath,graphicx}
\usepackage{color}
\usepackage{nameref}
\usepackage{hyperref}
\usepackage{soul}
\usepackage{color}
\usepackage{listings}
\usepackage[symbol]{footmisc}
\usepackage{dsfont}
\usepackage{amsmath}
\usepackage{mathtools}
\usepackage{centernot}
\usepackage{multirow}
\usepackage{makecell}
\usepackage{bm}
\usepackage{amsfonts}
\usepackage{adjustbox}
\usepackage{arydshln}
\usepackage{booktabs}       % professional-quality tables
\usepackage{xpatch}
\let\oldbibitem\bibitem

\newcommand{\newfootnotesize}{\fontsize{9pt}{9pt}\selectfont}

\renewcommand{\bibitem}[1]{\oldbibitem{#1}\newfootnotesize}
\newcommand*{\ditto}{---\texttt{"}---}

\usepackage{url}
\newcommand{\footurl}[1]{\begingroup\urlstyle{same}\url{#1}\endgroup}

\DeclarePairedDelimiter\floor{\lfloor}{\rfloor}

\DeclarePairedDelimiter\abs{\lvert}{\rvert}%
\makeatletter
\let\oldabs\abs
\def\abs{\@ifstar{\oldabs}{\oldabs*}}

\newcommand{\parn}[1]{\left(#1\right)}

\newlength{\NOTskip}

\title{HiFTNet: A Fast High-Quality Neural Vocoder with Harmonic-plus-Noise Filter and Inverse Short Time Fourier Transform}
\name{Yinghao Aaron Li, Cong Han, Xilin Jiang, Nima Mesgarani}
\address{Department of Electrical Engineering, Columbia University, USA}
\begin{document}
\ninept
\maketitle
\begin{abstract}
Recent advancements in speech synthesis have leveraged GAN-based networks like HiFi-GAN and BigVGAN to produce high-fidelity waveforms from mel-spectrograms. However, these networks are computationally expensive and parameter-heavy. iSTFTNet addresses these limitations by integrating inverse short-time Fourier transform (iSTFT) into the network, achieving both speed and parameter efficiency. In this paper, we introduce an extension to iSTFTNet, termed HiFTNet, which incorporates a harmonic-plus-noise source filter in the time-frequency domain that uses a sinusoidal source from the fundamental frequency (F0) inferred via a pre-trained F0 estimation network for fast inference speed. Subjective evaluations on LJSpeech show that our model significantly outperforms both iSTFTNet and HiFi-GAN, achieving ground-truth-level performance. HiFTNet also outperforms BigVGAN-base on LibriTTS for unseen speakers and achieves comparable performance to BigVGAN while being four times faster with only $1/6$ of the parameters. Our work sets a new benchmark for efficient, high-quality neural vocoding, paving the way for real-time applications that demand high quality speech synthesis.
\end{abstract}
\begin{keywords}
Waveform synthesis, mel-spectrogram vocoder, harmonic-plus-noise neural source filter, inverse short-time Fourier transform, generative adversarial networks
\end{keywords}
\section{Introduction}
\label{sec:intro}

Waveform synthesis plays a crucial role in modern speech generation technologies such as text-to-speech (TTS) and voice conversion (VC). These systems often employ a two-stage strategy: the first stage generates an intermediate representation, and the second stage converts it into waveforms. Mel-spectrograms have long been the favored intermediate representations in TTS \cite{shen2018natural, ren2020fastspeech, ren2019fastspeech, lancucki2021fastpitch, li2022styletts} and VC \cite{qian2019autovc, kaneko2020cyclegan, li2021starganv2, levkovitch2022zero, li2023stylettsvc, li2023slmgan} due to their closeness to human perceptions and reduced dimensionality. A vocoder that performs this second stage must infer missing phase information from the mel-spectrogram to reconstruct the waveform. The most effective and efficient methods so far have been adversarial generative networks (GAN) with convolutional neural network (CNN) architectures \cite{kumar2019melgan, yamamoto2020parallel, kong2020hifi, jang2021univnet, lee2022bigvgan}. While models like BigVGAN \cite{lee2022bigvgan} have obtained state-of-the-art performance in terms of synthesis quality, they are burdened by a large number of parameters required to generate waveforms directly from input mel-spectrograms, which hinders their application in real-time scenarios like TTS and VC. Therefore, the development of faster and more lightweight high-quality vocoders without sacrificing performance has become a pressing need.

In this paper, we introduce \textbf{H}armonics-plus-noise \textbf{i}nverse \textbf{F}ourier \textbf{T}ransform \textbf{Net}work (HiFTNet), a neural vocoder designed to meet these criteria. HiFTNet builds upon iSTFTNet \cite{kaneko2022istftnet} but goes beyond it to achieve high-quality waveform synthesis. Unlike previous vocoder models that generate waveform directly, HiFTNet follows iSTFTNet by modeling the magnitude and phase of the spectrogram and uses inverse short-time Fourier transform (iSTFT) for the final waveform generation. A key innovation in HiFTNet is its integration of a neural harmonic-plus-noise source filter \cite{wang2019neural} in the time-frequency domain using a sine wave source computed from the fundamental frequency (F0) extracted by a pre-trained F0 estimation network as opposed to traditional acoustic algorithms \cite{morise2009fast, morise2017harvest}. This modification substantially enhances the quality of the synthesized speech while minimally affecting the inference speed. 

Our evaluations demonstrate that HiFTNet significantly outperforms iSTFTNet and HiFi-GAN while maintaining similarly fast inference speed, achieving ground-truth level performance on LJSpeech \cite{ito2017lj} with a comparative mean opinion score (CMOS) of $-0.06$ ($p \gg 0.05$). Additionally, it is on par with BigVGAN on the LibriTTS \cite{zen2019libritts} dataset $\left(\text{CMOS} = 0.01, p \gg 0.05\right)$ but is $4\times$ faster and requires only $1/6$ of the parameters, thereby setting a new benchmark for efficient, high-quality neural vocoding. The demo samples are available at \url{https://hiftnet.github.io/}. 

\section{Methods}

\begin{figure}[!th]
    \centering
    \includegraphics[width=\columnwidth]{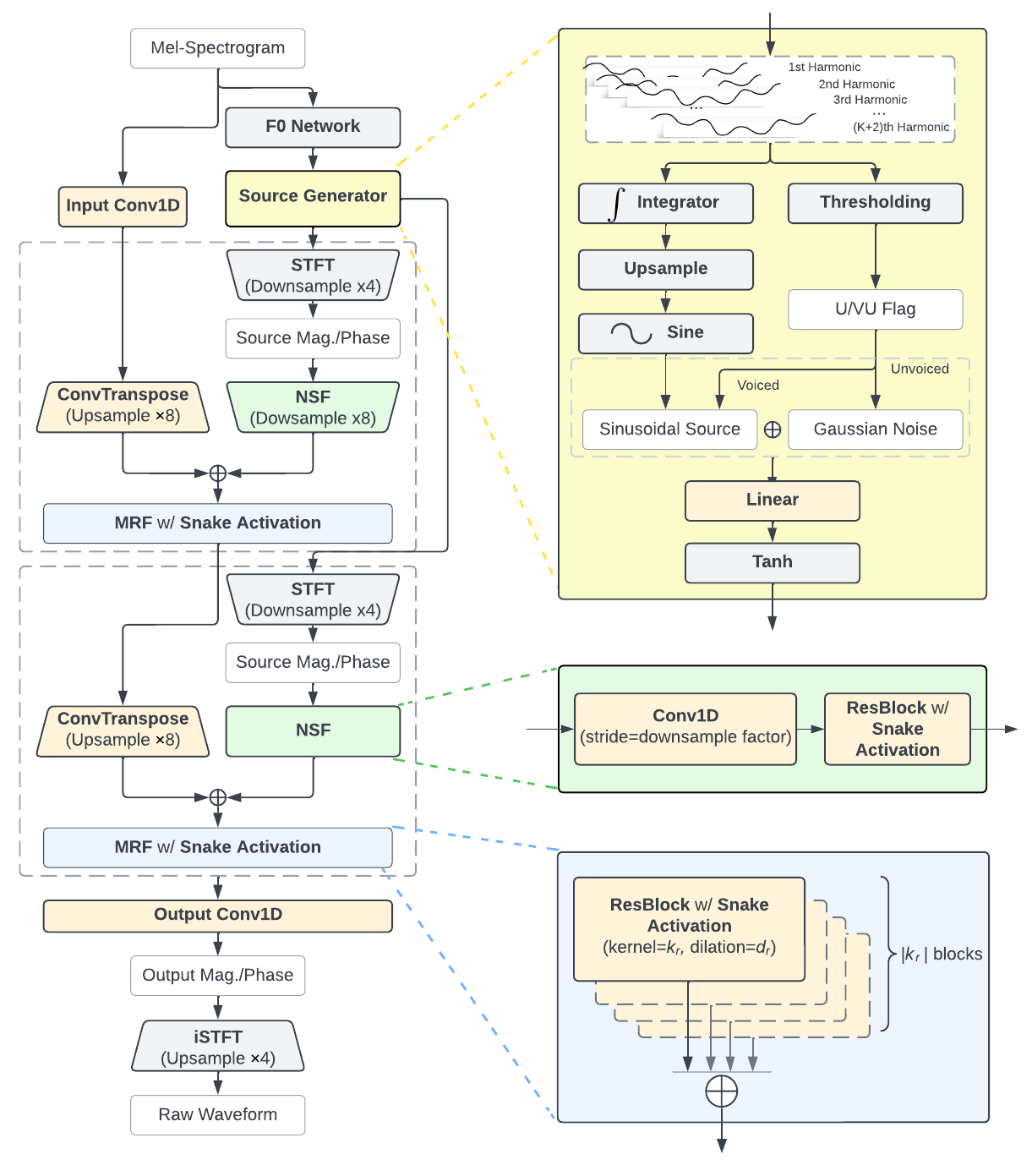}
    \caption{Overview of the HiFTNet architecture. The figure shows an example architecture of HiFTNet for 22.5 kHz audio generation with a hop size of 256. Orange modules are basic neural network components with tunable parameters during training, while grey modules are either pre-trained and fixed or non-trainable. The MRF module is the same as in HiFi-GAN \cite{kong2020hifi} that adds features from $|k_r|$ blocks but consists of ResBlocks with Snake functions instead of leaky ReLU. } 
    \label{fig:enter-label}
\end{figure}

\label{sec:hifnet}
HiFTNet builds upon the iSTFTNet \textit{V1-C8C8I} architecture \cite{kaneko2022istftnet} but introduces several key modifications. Firstly, we integrate a neural harmonic-plus-noise source filter \cite{wang2019neural} in the time-frequency domain, using the fundamental frequency extracted from the input mel-spectrogram via a pre-trained F0 network. We also substitute the MSD discriminator \cite{kong2020hifi} with the MRD discriminator \cite{jang2021univnet} and replace the leaky ReLU activation function in the generator with the Snake activation function \cite{ziyin2020neural}. Lastly, we adopt the truncated pointwise relativistic loss function \cite{li2022improve} to further enhance sound quality. The following sections elaborate on each of these modifications.

\subsection{Time-Frequency Harmonic-plus-Noise Source Filter}

Neural harmonic-plus-noise source filters (hn-NSF) \cite{wang2019neural} have found various applications in speech synthesis \cite{yoneyama2022unified, choi2022nansy++} and singing synthesis \cite{chen2020hifisinger, liu2022diffsinger}. These filters enhance the quality of the synthesized waveform by mitigating phase distortion. Generally, a sinusoidal source aligned in-phase with the target waveform is generated from its fundamental frequency (F0) for the voiced portions, while Gaussian noise fills the unvoiced segments. This source is then processed through a series of neural network layers named NSF. Here, we introduce several adjustments to better suit the iSTFTNet architecture and optimize inference speed, as detailed below.

\subsubsection{Efficient Source Generation}
We adopt the original hn-NSF source generation scheme %\footnote{An open-source implementation is available at \url{https://github.com/nii-yamagishilab/project-NN-Pytorch-scripts/tree/master/project/01-nsf}} 
presented in \cite{wang2019neural}, but with a critical change to significantly boost inference speed. In the original work \cite{wang2019neural}, the input fundamental frequency (F0) $p$ is initially upsampled to align with the sampling rate of the target waveform. It is then multiplied by a factor $i \in \{1, \ldots, K + 2\}$ to produce harmonic overtones $h_i$, where $K$ is the total number of harmonic overtones. Each $h_i$ is  integrated to yield the instantaneous phase $\varphi_i$ in radian for generating the sinusoidal source $s_i$: 

\begin{equation}
\label{eq1}
    h_i(t) = i \cdot p\left[\floor*{t \cdot f_s / L}  \right],
\end{equation}
\begin{equation}
\label{eq2}
    \varphi_i(t) = \parn{\frac{1}{f_s} \text{ mod } 1}\int_0^t h_i(t) \,dt,
\end{equation}
\begin{equation}
\label{eq3}
    s_i(t) = A\cdot\sin(2\pi \varphi_i(t)),
\end{equation}
\noindent where $f_s$ denotes the sampling rate of the target waveform, $L$ is the hop size, and $A$ is the source amplitude. It is worth noting that the integration operation is typically implemented via cumulative sum. Since $p$ originates from the mel-spectrogram domain, its length $N$ is considerably smaller than the target waveform length $T$. In Equation \ref{eq1}, $p$ is upsampled from size $N$ to $T$ where $T = NL$. The integration in Equation \ref{eq2} is of order $O(T)$, which greatly hinders the inference speed when dealing with long target waveforms. However, given that both upsampling and integration are linear operations, we can swap their order to reduce the complexity to $O(N)$:

\begin{equation}
\label{eq4}
    h_i[n] = i \cdot p[n],
\end{equation}
\begin{equation}
\label{eq5}
    \phi[n] = \parn{\frac{1}{f_s} \text{ mod } 1}\sum\limits_{k=0}^{n - 1} h_i[k],
\end{equation}
\begin{equation}
\label{eq6}
    \tilde{\varphi}(t) = L\cdot\phi\left[\floor*{t \cdot f_s / L}  \right],
\end{equation}
\begin{equation}
\label{eq7}
    s_i(t) = A\cdot\sin(2\pi \tilde{\varphi}_i(t)),
\end{equation}
where $\phi[n]$ is the instantaneous phase before upsampling and $\tilde{\varphi}_i(t) \approx \varphi_i(t)$\footnote{Although the upsampling and the continuous version of integration are both linear and can commute with each other, with the discrete version $\tilde{\varphi}_i(t) \neq \varphi_i(t)$ even after scaled by the hop size $L$. The difference between $\tilde{\varphi}_i(t)$ and $\varphi_i(t)$ is $\left[a_1, a_2, \ldots, a_N\right]$, where $a_i = \left[\parn{L-1}\cdot \phi[i], \parn{L-2}\cdot\phi[i], \ldots, \parn{L-L}\cdot\phi[i]\right]$ is the adjusting factor of length $L$. However, since this additional adjusting factor does not add new information to the neural source filter, we noticed that there is no difference in sound quality regardless of whether the adjusting factor is subtracted from ${\tilde{\varphi}(t)}$. }. We note that the $L$ factor in Equation \ref{eq6} scales the value by the hop size, as Equation \ref{eq5} now integrates with $1/L$ of steps compared to steps in Equation \ref{eq2}.

Gaussian noise serves as the source for the unvoiced segments. An unvoiced (UV) flag is set by applying a 10 Hz threshold to the input F0, marking frames with F0 values below this as unvoiced. The final excitation source for the $i$th harmonic overtone is expressed as:
\begin{equation}
    x_i(t) = (1 - UV(t))s_i(t) + UV(t)\xi,
\end{equation}
where $\xi \sim \mathcal{N}(0, A/3)$. Following \cite{wang2019neural}, we set $A = 0.1$. \\
Finally, all harmonics are linearly combined and processed through a tanh function, as shown in the yellow block of Figure \ref{fig:enter-label}:
\begin{equation}
    x(t) = \text{tanh}\parn{\sum\limits_{i=1}^{K+2} w_i x_i(t)},
\end{equation}
where $w_i$ are learnable parameters and $K = 8$ following \cite{liu2022diffsinger}. 

\subsubsection{F0 Estimation with Pre-Trained Neural Network}
In both the original hn-NSF model \cite{wang2019neural} and subsequent vocoder works \cite{yoneyama2022unified}, the F0 for source generation is derived using the WORLD vocoder \cite{morise2016world}. However, as shown in our prior research \cite{li2022styletts}, traditional acoustic algorithms \cite{ morise2009fast, morise2017harvest} for pitch extraction tend to be both inaccurate and failure-prone, negatively affecting reconstruction quality. Furthermore, commonly applied algorithms for pitch extraction, such as distributed inline-filter operation (DIO) \cite{morise2009fast} and Harvest \cite{morise2017harvest}, have an $O(N\log N)$ complexity and run on the CPU without GPU acceleration. Most critically, these algorithms operate in the time domain, requiring the very waveform input we aim to synthesize from mel-spectrograms for F0 extraction.

To address these limitations, we employ a neural network for F0 estimation. Specifically, we adopt the approach in \cite{li2021starganv2} that pre-trains a JDC network \cite{kum2019joint} using pitch labels extracted with DIO and Harvest, supplemented with standard data augmentation techniques in speech recognition \cite{ko2015audi}. This pre-trained network is then used for more accurate and robust F0 estimation from the input mel-spectrograms. Performance with alternative architecture without the LSTM RNN in the JDC network is also explored in section \ref{sec:ab}.

\subsubsection{Time-Frequency Neural Source Filter}

In HiFTNet, the final output of the generator consists of the magnitude and phase of the spectrogram rather than waveforms. Consequently, the neural source filters must also process the excitation source within the time-frequency domain to align with this output. Instead of directly feeding the source waveforms to the neural source filter (NSF) module, we initially perform an STFT transformation using the same parameters (FFT size, hop size, and window length) as the terminating inverse STFT operation in the network output, thereby converting the source waveform to the time-frequency domain. Section \ref{sec:ab} demonstrates that this time-frequency processing is crucial for high-quality waveform synthesis, as substituting the STFT module with a learnable CNN module of the same stride as the hop size and the same number of output channels as the FTT size significantly deteriorates performance.

In contrast to the complex NSF modules described in \cite{wang2019neural}, our NSF module is only composed of a 1D convolutional layer for source downsampling to match the intermediate feature size, followed by a residual block for fast inference, as illustrated in Figure \ref{fig:enter-label}. We find that this architecture suffices for generating high-quality samples.

\subsection{MRD Discriminator and Snake Function}

We substitute the original multi-scale discriminator (MSD) from iSTFTNet with the multi-resolution discriminator (MRD) as introduced in \cite{jang2021univnet}. This change has been demonstrated to enhance sound quality in subsequent studies \cite{lee2022bigvgan}. We retain the multi-period discriminator (MPD) initially proposed in \cite{kong2020hifi}, applying the same LSGAN \cite{mao2017least} objective for both generator and discriminator training. Additionally, we employ the same feature matching loss during the generator training as in \cite{kong2020hifi}, a technique commonly adopted in contemporary neural vocoders \cite{jang2021univnet, lee2022bigvgan, kaneko2022istftnet}.

Furthermore, we replace leaky ReLU activation functions across the generator with  Snake functions \cite{ziyin2020neural}, first proposed for speech synthesis in BigVGAN \cite{lee2022bigvgan}. The Snake function is defined as:

\begin{equation}
    f_\alpha(x) = x + \frac{1}{\alpha}\sin^2(\alpha x),
\end{equation}
\noindent where $\alpha$ is a learnable parameter. Although the generator's final output is not a waveform but rather the magnitude and phase of the spectrogram, these are still highly periodic, especially the phase. As such, employing the Snake activation function aids in the model's capacity to learn the periodic structure of the speech signal. This is also in line with what we have found in our previous work \cite{li2023styletts} where iSTFTNet is used as the speech decoder for human-level TTS. Unlike BigVGAN \cite{lee2022bigvgan}, we do not include the anti-aliasing filter for upsampling. This is primarily due to the instability introduced by the filter, and also because our generator consists of only two upsampling modules, resulting in less aliasing compared to previous vocoders that synthesize waveforms directly.

\subsection{Truncated Pointwise Relativistic Loss Function}

To further enhance sound quality during adversarial training, we incorporate the Truncated Pointwise Relativistic (TPR) loss function \cite{li2022improve}. This approach has proven successful in our previous work for achieving human-level TTS with iSTFTNet-based decoders \cite{li2023styletts}. This loss function aims to quantify the disparity between the discriminator's outputs for the real target waveform 
$\bm{y}$ and the generated or reconstructed waveform $\bm{\hat{y}}$. Specifically, the TPR loss encourages the discriminator to assign lower scores to the generated samples relative to their real counterparts for each frame point. Conversely, it motivates the generator to produce samples that the discriminator would rate higher compared to the real samples for each frame point.

The loss is formulated using the relativistic difference $ \mathcal{R}(\bm{{y}}, \bm{\hat{y}})$:
\begin{equation}
    \mathcal{R}(\bm{{y}}, \bm{\hat{y}}) = D(\bm{y}) - D\parn{\hat{\bm{y}}} - m(\bm{y}, \bm{\hat{y}}),
\end{equation}
\begin{equation}
      m(\bm{y}, \bm{\hat{y}}) = \mathbb{M}_{\bm{y}, \bm{\hat{y}}}\left[D({\bm{y}}) - D(\hat{\bm{y}})\right]. \\
\end{equation}
Here, $D(\cdot)$ denotes both MPD and MRD outputs, and $m(\bm{y}, \bm{\hat{y}})$ is the median of the relativistic difference in a batch, calculated via $\mathbb{M}\left[\cdot\right]$, the median operation. The TPR loss is thus defined as:
\begin{equation}
      \mathcal{L}_\text{rel}(D;G) = \tau - {\text{ } \mathbb{E}_{ \{\mathcal{R}(\bm{{y}}, \bm{\hat{y}}) \leq 0 \}}\left[\text{ReLU}\parn{\tau - {\mathcal{R}(\bm{{y}}, \bm{\hat{y}})^2 }} \right]}, \\
\end{equation}
\begin{equation}
      \mathcal{L}_\text{rel}(G;D) = \tau - {\text{ } \mathbb{E}_{ \{\mathcal{R}(\bm{\hat{y}}, \bm{y}) \leq 0 \}}\left[\text{ReLU}\parn{\tau - {\mathcal{R}(\bm{\hat{y}}, \bm{y})^2 }} \right]}, \\
\end{equation}
\noindent where $\{\mathcal{R}(\bm{{y}}, \bm{\hat{y}}) \leq 0 \}$ and $ \{\mathcal{R}(\bm{\hat{y}}, \bm{y}) \leq 0 \}$  denote the sets of $\bm{y}$ and $\bm{\hat{y}}$ that satisfy the respective conditions in a batch, $\text{ReLU}(\cdot) = \max(\cdot, 0)$, and $\tau$ is the truncation factor, set to 0.04 per \cite{li2022improve}. 

\section{Experiments}
\subsection{Datasets, Models and Training Details}
We conducted evaluations using the LJSpeech \cite{ito2017lj} and LibriTTS \cite{zen2019libritts} datasets. The LJSpeech dataset, which comprises 13,100 short audio clips totaling approximately 24 hours, was used for training our single-speaker model. We compared this model to HiFi-GAN and iSTFTNet, both also trained on the LJSpeech dataset. The dataset was partitioned into 12,950 training and 150 validation samples, following the same split used in \cite{kong2020hifi}. For our multi-speaker model, we employed the combined LibriTTS \textit{train-960} subset \cite{zen2019libritts}, which is sourced from \textit{train-clean-100}, \textit{train-clean-360}, and \textit{train-other-500} subsets per \cite{lee2022bigvgan}. This dataset contains around 555 hours of audio from 2,311 speakers. We compared our model to BigVGAN-base and BigVGAN on the \textit{test-clean} and \textit{test-other} subsets for unseen speakers. The former subset comprises clean speech, while the latter contains noisier samples.

We followed the pre-processing pipeline of 22.5 kHz audio as in \cite{kong2020hifi} for generating the mel-spectrograms. Specifically, we used a hop size of 256, an FFT size of 1024, a window length of 1024, a lowest frequency of 0 Hz, and the highest frequency of 8000 Hz with 80 mel bins. Audio samples from the LibriTTS dataset were downsampled to 22.5 kHz to align with this pre-processing. Our model was trained for 500k steps on both the LJSpeech and LibriTTS datasets, with a batch size of 16 one-second-long audio segments on a single NVIDIA A40 GPU. We employed the AdamW optimizer \cite{loshchilov2018fixing} with 
 $\beta_1 = 0.8, \beta_2 = 0.99$, weight decay $\lambda = 0.01$, and an initial learning rate $\gamma = 0.0002$ with an 
 exponential decay rate of $0.999$. 

For comparison, we used official pre-trained checkpoints for HiFi-GAN on LJSpeech \footnote{Available at \url{https://github.com/jik876/hifi-gan}} and BigVGAN on LibriTTS \footnote{Available at \url{https://github.com/NVIDIA/BigVGAN}}. As there was no official iSTFTNet implementation and checkpoint, we trained an iSTFTNet baseline model using the same hyperparameters with an unofficial implementation \footnote{ \url{https://github.com/rishikksh20/iSTFTNet-pytorch}} for 500k steps.

\subsection{Evaluations}
To assess model performance, we employed both subjective and objective evaluation methods. For the subjective assessments, we used the Comparative Mean Opinion Score (CMOS) metric to establish statistical significance as the differences between these models are subtle and not readily noticeable. This allows raters to discern subtle differences often overlooked in traditional MOS experiments \cite{li2023styletts}. We recruited native English speakers located in the U.S. via Amazon Mechanical Turk for these evaluations. Participants were guided to listen to paired samples from distinct models using headphones and then rate the second sample as better or worse than the first, using a scale from -6 to 6 in increments of 1. Each test comprised 30 randomly selected audio samples from the test dataset, which were converted into mel-spectrograms and then back into waveforms using both our model and the baseline models. We also included three attention-checker pairs containing identical audio clips. Raters who assigned these pairs an average score more than $\pm 0.5$ were excluded from the results. Each evaluation set involved a minimum of ten raters, ensuring at least five had passed the attention checks.

For objective evaluations, we relied on mel-cepstral distortion (MCD) with dynamic time warping calculated using an open source implementation \footnote{ \url{https://github.com/chenqi008/pymcd/}} as a metric to compare the synthesized waveform with the ground-truth audio. To assess inference speed, we computed the real-time factor (RTF) using an NVIDIA RTX 3090 Ti GPU.

\section{Results}
\subsection{Model Performance}
\begin{table}[!t]
\label{tab:1}
  \caption{Comparative mean opinion scores (CMOS) for HiFTNet with p-values from Wilcoxon test relative to other models, mel-spectral distortion (MCD) relative to ground truth, inference speed (RTF), and GPU RAM usage when synthesizing a 10-second audio. For CMOS, positive scores indicate that HiFTNet is better. For the dataset column, \textit{test-clean} and \textit{test-other} represent the results on the corresponding subsets of the LibriTTS dataset.}
  \label{tab:1}
  \centering
\begin{adjustbox}{width=\columnwidth,center}
  \begin{tabular}{llccccc}
    \toprule
    \textbf{Model} & \textbf{Dataset} & \textbf{CMOS  (p-value)} $\uparrow$ & \textbf{MCD} $\downarrow$ & \textbf{RTF} $\downarrow$ & \textbf{RAM} $\downarrow$ \\ 
    \midrule
    Ground Truth  & LJSpeech   &  ${-}${0.06} ({$p = 0.396$}) & --- & --- & --- \\
    HiFTNet & LJSpeech & --- & 2.567 & 0.0057 & 0.90 GB\\
    iSTFTNet   & LJSpeech   &  {${+}${0.64}} ($p < 10^{-7}$) & 2.820 & 0.0031 & 0.77 GB \\
    HiFi-GAN   & LJSpeech   &  {${+}${0.19}} ($p = 0.028$) & 2.816 & 0.0043 & 0.75 GB \\
    \midrule
    Ground Truth  & \textit{test-clean}  &  ${-}${0.21}  $(p = 0.033)$ & --- & --- & --- \\
    HiFTNet & \textit{test-clean} & --- & 2.892 & \ditto & \ditto\\

    BigVGAN-base  & \textit{test-clean}  &  {${+}${0.21}}  $(p = 0.001)$ & 3.079 & 0.0159 & 0.90 GB \\
    BigVGAN  & \textit{test-clean}  &  ${-}${0.05}  $(p = 0.552)$ & 2.656 & 0.0243 & 1.52 GB \\
    \midrule
    Ground Truth  & \textit{test-other}  &  ${-}${0.10}  $(p = 0.189)$ & --- & --- & --- \\
    HiFTNet & \textit{test-other} & --- & 3.690 & \ditto & \ditto\\
    BigVGAN-base  & \textit{test-other}  &  {${+}${0.17}}  $(p = 0.022)$ & 3.892 & \ditto & \ditto \\
    BigVGAN  & \textit{test-other}  &  ${+}${0.12}  $(p = 0.354)$ & 3.189 & \ditto & \ditto \\
    \bottomrule

  \end{tabular}
  \end{adjustbox}
\vspace{-10 pt}
\end{table}

As illustrated in Table \ref{tab:1}, HiFTNet exhibits a CMOS score of -0.06 with $p \gg 0.05$ when tested on the LJSpeech dataset. This essentially places our model on par with the ground truth for this particular dataset. Moreover, HiFTNet has significantly outperformed both iSTFTNet and HiFi-GAN in terms of CMOS ($p < 0.05$) and MCD, while incurring only a minor increase in inference speed and RAM.

When evaluated on the LibriTTS \textit{test-clean} subset, HiFTNet significantly surpasses BigVGAN, with a CMOS of 0.21 ($p < 0.05$) and also a slightly lower MCD. This is achieved while maintaining the same RAM usage yet being $2.5 \times$ faster. Furthermore, HiFTNet demonstrates performance comparable to BigVGAN with a CMOS of  $-0.05 $ $(p \gg 0.05)$, but operates $4 \times$ faster and consumes only half the GPU RAM during inference. Similar trends are observed on the \textit{test-other} dataset, where HiFTNet notably outperforms BigVGAN-base and achieves performance akin to BigVGAN. 

Together, HiFTNet achieves a CMOS of 0.013 $(p = 0.873)$ compared to BigVGAN on the LibriTTS dataset for unseen speakers. Notably, HiFTNet accomplishes all this with only 17.7 M trainable parameters, making it approximately 1/6 lighter in size compared to BigVGAN's 114 M parameters. This positions HiFTNet as a viable alternative to BigVGAN in end-to-end training scenarios, such as speech language model (SLM) adversarial training with SLM feature matching loss in our recently proposed VC model \cite{li2023slmgan}, thanks to its more efficient RAM usage and faster inference speed.

\subsection{Ablation Study}
\label{sec:ab}
\begin{table}[!t]
\label{tab:2}
  \caption{CMOS of proposed model relative to component-ablated models, MCD relative to ground truth, and inference speed (RTF).  }
  \label{tab:1}
  \centering
  \begin{tabular}{lcccc}
    \toprule
    \textbf{Model}  & \textbf{CMOS} $\uparrow$ & \textbf{MCD} $\downarrow$  & \textbf{RTF} $\downarrow$   \\ 
    \midrule
    Baseline     &  \textbf{0} & \textbf{2.567} & 0.0057  \\
    w/o hn-NSF     &  ${-}$1.116 & 2.929 & \textbf{0.0036}  \\
    w/o STFT  & ${-}$0.358 &  2.716  & 0.0055 \\
    w/o Snake     &  ${-}$0.108 & 2.689 & 0.0050  \\
    w/o LSTM     &  ${-}$0.475 & 2.639 & 0.0047  \\
    \bottomrule

  \end{tabular}
\vspace{-10 pt}
\end{table}

In table \ref{tab:2}, we present the CMOS of the proposed model compared to models with components ablated to demonstrate the effectiveness of our proposed components. Omitting the hn-NSF results in a dramatic performance decline, reflected by a CMOS of $-1.116$, making the model inferior to iSTFTNet. Substituting the STFT modules with trainable 1D convolutional layers prior to the NSF also yields a reduced CMOS of $-0.358$. Additionally, switching the Snake activation function back to leaky ReLU causes a minor performance dip, indicated by a CMOS of $-0.108$. Finally, removing the LSTM layer from the pitch extraction network, while accelerating inference time, significantly degrades performance with a CMOS of $-0.475$.

These findings affirm the efficacy of each proposed component in enhancing performance, although some may slightly increase inference time. The Snake activation function, for example, decelerates the system by approximately 15\% but only marginally bolsters performance, making it an optional component if inference speed is paramount. Intriguingly, removing the LSTM from the F0 extraction network has a negative impact on performance, implying that F0 estimation quality is a critical factor for vocoder performance. This suggests that, even though F0 is largely a local feature, some global information not captured by CNN still contributes to accurate F0 estimation needed for high-quality speech synthesis.

\section{Conclusions}
In this study, we introduced HiFTNet, a neural vocoder model that
offers substantial improvements in sound quality and inference speed
over existing models like iSTFTNet, HiFi-GAN and BigVGAN-base, with performance comparable to significantly larger models
such as BigVGAN. Leveraging a suite of novel components, including the time-frequency harmonic-plus-noise neural source filter, the
Snake activation function, and a MRD discriminator and TPR loss,
our model achieved superior performance across multiple metrics
and datasets. The ablation study further corroborated the importance
of each component, highlighting their individual contributions to the
model’s efficacy. The study also suggests a future research direction
in optimizing neural networks for faster and more precise F0 estimation to further enhance the performance and inference speed of
hn-NSF-based vocoders.

\section{Acknowledgements}
This work was funded by the National Institutes of Health (NIH-NIDCD) and a grant from Marie-Josee and Henry R. Kravis. %We used ChatGPT to improve the readability of the writing.


\newpage

\bibliographystyle{IEEEbib}
\bibliography{refs}

\begin{thebibliography}{10}

\bibitem{shen2018natural}
Jonathan Shen, Ruoming Pang, Ron~J Weiss, Mike Schuster, Navdeep Jaitly,
  Zongheng Yang, Zhifeng Chen, Yu~Zhang, Yuxuan Wang, Rj~Skerrv-Ryan, et~al.,
\newblock ``Natural tts synthesis by conditioning wavenet on mel spectrogram
  predictions,''
\newblock in {\em 2018 IEEE international conference on acoustics, speech and
  signal processing (ICASSP)}. IEEE, 2018, pp. 4779--4783.

\bibitem{ren2020fastspeech}
Yi~Ren, Chenxu Hu, Xu~Tan, Tao Qin, Sheng Zhao, Zhou Zhao, and Tie-Yan Liu,
\newblock ``Fastspeech 2: Fast and high-quality end-to-end text to speech,''
\newblock {\em arXiv preprint arXiv:2006.04558}, 2020.

\bibitem{ren2019fastspeech}
Yi~Ren, Yangjun Ruan, Xu~Tan, Tao Qin, Sheng Zhao, Zhou Zhao, and Tie-Yan Liu,
\newblock ``Fastspeech: Fast, robust and controllable text to speech,''
\newblock {\em Advances in neural information processing systems}, vol. 32,
  2019.

\bibitem{lancucki2021fastpitch}
Adrian {\L}a{\'n}cucki,
\newblock ``Fastpitch: Parallel text-to-speech with pitch prediction,''
\newblock in {\em ICASSP 2021-2021 IEEE International Conference on Acoustics,
  Speech and Signal Processing (ICASSP)}. IEEE, 2021, pp. 6588--6592.

\bibitem{li2022styletts}
Yinghao~Aaron Li, Cong Han, and Nima Mesgarani,
\newblock ``Styletts: A style-based generative model for natural and diverse
  text-to-speech synthesis,''
\newblock {\em arXiv preprint arXiv:2205.15439}, 2022.

\bibitem{qian2019autovc}
Kaizhi Qian, Yang Zhang, Shiyu Chang, Xuesong Yang, and Mark Hasegawa-Johnson,
\newblock ``Autovc: Zero-shot voice style transfer with only autoencoder
  loss,''
\newblock in {\em International Conference on Machine Learning}. PMLR, 2019,
  pp. 5210--5219.

\bibitem{kaneko2020cyclegan}
Takuhiro Kaneko, Hirokazu Kameoka, Kou Tanaka, and Nobukatsu Hojo,
\newblock ``Cyclegan-vc3: Examining and improving cyclegan-vcs for
  mel-spectrogram conversion,''
\newblock {\em arXiv preprint arXiv:2010.11672}, 2020.

\bibitem{li2021starganv2}
Yinghao~Aaron Li, Ali Zare, and Nima Mesgarani,
\newblock ``Starganv2-vc: A diverse, unsupervised, non-parallel framework for
  natural-sounding voice conversion,''
\newblock {\em arXiv preprint arXiv:2107.10394}, 2021.

\bibitem{levkovitch2022zero}
Alon Levkovitch, Eliya Nachmani, and Lior Wolf,
\newblock ``Zero-shot voice conditioning for denoising diffusion tts models,''
\newblock {\em arXiv preprint arXiv:2206.02246}, 2022.

\bibitem{li2023stylettsvc}
Yinghao~Aaron Li, Cong Han, and Nima Mesgarani,
\newblock ``Styletts-vc: One-shot voice conversion by knowledge transfer from
  style-based tts models,''
\newblock in {\em 2022 IEEE Spoken Language Technology Workshop (SLT)}. IEEE,
  2023, pp. 920--927.

\bibitem{li2023slmgan}
Yinghao~Aaron Li, Cong Han, and Nima Mesgarani,
\newblock ``Slmgan: Exploiting speech language model representations for
  unsupervised zero-shot voice conversion in gans,''
\newblock {\em arXiv preprint arXiv:2307.09435}, 2023.

\bibitem{kumar2019melgan}
Kundan Kumar, Rithesh Kumar, Thibault De~Boissiere, Lucas Gestin, Wei~Zhen
  Teoh, Jose Sotelo, Alexandre De~Brebisson, Yoshua Bengio, and Aaron~C
  Courville,
\newblock ``Melgan: Generative adversarial networks for conditional waveform
  synthesis,''
\newblock {\em Advances in neural information processing systems}, vol. 32,
  2019.

\bibitem{yamamoto2020parallel}
Ryuichi Yamamoto, Eunwoo Song, and Jae-Min Kim,
\newblock ``Parallel wavegan: A fast waveform generation model based on
  generative adversarial networks with multi-resolution spectrogram,''
\newblock in {\em ICASSP 2020-2020 IEEE International Conference on Acoustics,
  Speech and Signal Processing (ICASSP)}. IEEE, 2020, pp. 6199--6203.

\bibitem{kong2020hifi}
Jungil Kong, Jaehyeon Kim, and Jaekyoung Bae,
\newblock ``Hifi-gan: Generative adversarial networks for efficient and high
  fidelity speech synthesis,''
\newblock {\em Advances in Neural Information Processing Systems}, vol. 33, pp.
  17022--17033, 2020.

\bibitem{jang2021univnet}
Won Jang, Dan Lim, Jaesam Yoon, Bongwan Kim, and Juntae Kim,
\newblock ``Univnet: A neural vocoder with multi-resolution spectrogram
  discriminators for high-fidelity waveform generation,''
\newblock {\em arXiv preprint arXiv:2106.07889}, 2021.

\bibitem{lee2022bigvgan}
Sang-gil Lee, Wei Ping, Boris Ginsburg, Bryan Catanzaro, and Sungroh Yoon,
\newblock ``Bigvgan: A universal neural vocoder with large-scale training,''
\newblock {\em arXiv preprint arXiv:2206.04658}, 2022.

\bibitem{kaneko2022istftnet}
Takuhiro Kaneko, Kou Tanaka, Hirokazu Kameoka, and Shogo Seki,
\newblock ``istftnet: Fast and lightweight mel-spectrogram vocoder
  incorporating inverse short-time fourier transform,''
\newblock in {\em ICASSP 2022-2022 IEEE International Conference on Acoustics,
  Speech and Signal Processing (ICASSP)}. IEEE, 2022, pp. 6207--6211.

\bibitem{wang2019neural}
Xin Wang, Shinji Takaki, and Junichi Yamagishi,
\newblock ``Neural source-filter waveform models for statistical parametric
  speech synthesis,''
\newblock {\em IEEE/ACM Transactions on Audio, Speech, and Language
  Processing}, vol. 28, pp. 402--415, 2019.

\bibitem{morise2009fast}
Masanori Morise, Hideki Kawahara, and Haruhiro Katayose,
\newblock ``Fast and reliable f0 estimation method based on the period
  extraction of vocal fold vibration of singing voice and speech,''
\newblock in {\em Audio Engineering Society Conference: 35th International
  Conference: Audio for Games}. Audio Engineering Society, 2009.

\bibitem{morise2017harvest}
Masanori Morise et~al.,
\newblock ``Harvest: A high-performance fundamental frequency estimator from
  speech signals.,''
\newblock in {\em INTERSPEECH}, 2017, pp. 2321--2325.

\bibitem{ito2017lj}
Keith Ito and Linda Johnson,
\newblock ``The lj speech dataset,''
\newblock 2017.

\bibitem{zen2019libritts}
Heiga Zen, Viet Dang, Rob Clark, Yu~Zhang, Ron~J Weiss, Ye~Jia, Zhifeng Chen,
  and Yonghui Wu,
\newblock ``Libritts: A corpus derived from librispeech for text-to-speech,''
\newblock {\em arXiv preprint arXiv:1904.02882}, 2019.

\bibitem{ziyin2020neural}
Liu Ziyin, Tilman Hartwig, and Masahito Ueda,
\newblock ``Neural networks fail to learn periodic functions and how to fix
  it,''
\newblock {\em Advances in Neural Information Processing Systems}, vol. 33, pp.
  1583--1594, 2020.

\bibitem{li2022improve}
Yanli Li and Congyi Wang,
\newblock ``Improve gan-based neural vocoder using truncated pointwise
  relativistic least square gan,''
\newblock in {\em Proceedings of the 4th International Conference on Advanced
  Information Science and System}, 2022, pp. 1--7.

\bibitem{yoneyama2022unified}
Reo Yoneyama, Yi-Chiao Wu, and Tomoki Toda,
\newblock ``Unified source-filter gan with harmonic-plus-noise source
  excitation generation,''
\newblock {\em arXiv preprint arXiv:2205.06053}, 2022.

\bibitem{choi2022nansy++}
Hyeong-Seok Choi, Jinhyeok Yang, Juheon Lee, and Hyeongju Kim,
\newblock ``Nansy++: Unified voice synthesis with neural analysis and
  synthesis,''
\newblock {\em arXiv preprint arXiv:2211.09407}, 2022.

\bibitem{chen2020hifisinger}
Jiawei Chen, Xu~Tan, Jian Luan, Tao Qin, and Tie-Yan Liu,
\newblock ``Hifisinger: Towards high-fidelity neural singing voice synthesis,''
\newblock {\em arXiv preprint arXiv:2009.01776}, 2020.

\bibitem{liu2022diffsinger}
Jinglin Liu, Chengxi Li, Yi~Ren, Feiyang Chen, and Zhou Zhao,
\newblock ``Diffsinger: Singing voice synthesis via shallow diffusion
  mechanism,''
\newblock in {\em Proceedings of the AAAI conference on artificial
  intelligence}, 2022, vol.~36, pp. 11020--11028.

\bibitem{morise2016world}
Masanori Morise, Fumiya Yokomori, and Kenji Ozawa,
\newblock ``World: a vocoder-based high-quality speech synthesis system for
  real-time applications,''
\newblock {\em IEICE TRANSACTIONS on Information and Systems}, vol. 99, no. 7,
  pp. 1877--1884, 2016.

\bibitem{kum2019joint}
Sangeun Kum and Juhan Nam,
\newblock ``Joint detection and classification of singing voice melody using
  convolutional recurrent neural networks,''
\newblock {\em Applied Sciences}, vol. 9, no. 7, pp. 1324, 2019.

\bibitem{ko2015audi}
Tom Ko, Vijayaditya Peddinti, Daniel Povey, and Sanjeev Khudanpur,
\newblock ``Audio augmentation for speech recognition,''
\newblock in {\em Sixteenth annual conference of the international speech
  communication association}, 2015.

\bibitem{mao2017least}
Xudong Mao, Qing Li, Haoran Xie, Raymond~YK Lau, Zhen Wang, and Stephen
  Paul~Smolley,
\newblock ``Least squares generative adversarial networks,''
\newblock in {\em Proceedings of the IEEE international conference on computer
  vision}, 2017, pp. 2794--2802.

\bibitem{li2023styletts}
Yinghao~Aaron Li, Cong Han, Vinay~S Raghavan, Gavin Mischler, and Nima
  Mesgarani,
\newblock ``Styletts 2: Towards human-level text-to-speech through style
  diffusion and adversarial training with large speech language models,''
\newblock {\em arXiv preprint arXiv:2306.07691}, 2023.

\bibitem{loshchilov2018fixing}
Ilya Loshchilov and Frank Hutter,
\newblock ``Fixing weight decay regularization in {Adam},'' 2018.

\end{thebibliography}

\end{document}